\documentstyle[12pt,epsfig]{article} 
\begin{document} 
\bibliographystyle{unsrt} 
 \begin{center} 
 {\large\bf Distribution of Return Periods of Rare Events 
 in Correlated \\Time Series}
\vspace{1.0cm} 

C. Pennetta$^*$ and  E.Alfinito 

\vspace{0.5cm} 
{\it Dipartimento di Ingegneria dell'Innovazione, 
  Universit\`a di Lecce and \\National    
  Nanotechnology Laboratory, CNR - INFM, Via Arnesano, 73100 Lecce, Italy.
  \\$^*$corresponding author: cecilia.pennetta@unile.it}  
\end{center} 

\vspace{0.5cm} 

{\small{\bf keywords}: Extreme values in random processes, 
Fluctuation phenomena,Time series analysis} 

 

\vspace{0.5cm} 
    
\begin{abstract} 
We study the effect on the distribution of return periods of rare events 
of the presence in a time series of finite-term correlations with 
non-exponential decay. Precisely, we analyze the auto-correlation function
and the statistics of the return intervals of extreme values of the resistance
fluctuations displayed by a resistor with granular structure in a
nonequilibrium stationary state. The resistance fluctuations, $\delta R$, are 
calculated by Monte Carlo simulations using the SBRN model introduced 
some years ago by Pennetta, Tref\'an and Reggiani and based on a resistor 
network approach. A rare event occurs when $\delta R$ overcomes a threshold 
value $q$ significantly higher than the average value of the resistance. 
We have found that for highly disordered networks, when the auto-correlation 
function displays a non-exponential decay but yet the resistance fluctuations 
are characterized by a finite correlation time, the distribution of 
return intervals of the extreme values is well described by a stretched 
exponential, with exponent largely independent of the threshold $q$. We 
discuss this result and some of the main open questions related to it, 
also in connection with very recent findings by other authors concerning 
the observation of stretched exponential distributions of return 
intervals of extreme events in long-term correlated time series. 
\end{abstract}  
  
 
\vspace*{0.5cm}  
\section{Introduction}  
It is well known that in a Poisson process the return periods between 
extreme events associated with the exceedance of a threshold are exponentially 
distributed \cite{storch,kotz}. In other terms, by considering a time 
series $x(t)$ made by uncorrelated records, the return periods $r_q$ of 
events above a threshold $q$ (i.e. the time intervals between two consecutive 
occurences of the condition $x(t)>q$) are distributed according to the 
probability density function (PDF): 
\begin{equation}      
P_q(r)=(1/{R_q}) \exp (-r/{R_q})  \label{poisson} 
\end{equation}
where $R_q$ is the mean return interval of these events 
\cite{storch,kotz}. Of course, the higher the value of $q$ (quantile), the
rarer are the events over the threshold and bigger is the value of $R_q$.
Fluctuations of prices in financial markets, wind speed 
data or daily precipitations in a given location for a same time windows are 
typically described by uncorrelated records \cite{storch,kantz_2005}.    
However, particularly in the last ten years, it has become clear that several
other important examples of time series display long-term correlations  
\cite{kantz_2005,bunde_physa2003,bunde_prl2005}. This is the case of 
physiological data, like heartbeats \cite{bunde_prl2000,ashkenazy} and 
neuron spikes \cite{davidsen}, hydro-meteorological records, like daily 
temperatures \cite{kantz_2005,bunde_physa2003,koscielny_prl98}, 
geophysical or astrophysical data, concerning for example the occurrence 
of earthquakes \cite{bak,corral} or solar flares \cite{boffetta}, 
stock market volatility \cite{kantz_2005,liu} and internet traffic records 
\cite{bunde_physa2003}. Long-term correlated series are characterized by an 
auto-correlation function which decays as a power-law: 
\begin{equation}
C_x(s) = <x_ix_{i+s}> = { 1 \over N-s} \sum_{i=1}^{N-s} x_ix_{i+s}
\sim s^{-\gamma_x} \ \ ,\ \ \ 0<\gamma_x<1  \label{long_corr}  
\end{equation}  
In this case, since the mean correlation time $\tau$ is the integral of 
$C_x(s)$ over $s$, it is easy to see that this time diverges when the 
correlation exponent $\gamma_x$ is between 0 and 1. 

Recently, Bunde et al. \cite{bunde_physa2003} investigated the effect of 
long-terms correlations on the statistics of the return periods of extreme 
events $r_q$. These authors found that: i) long-terms correlations 
leave unchanged the mean return interval $R_q$; ii) they significantly 
modify the distribution of return intervals whose PDF becomes a stretched 
exponential
\begin{equation}  
P_q(r) = a \ \exp\bigl[ -\bigl(b \ \ r / R_q \bigr)^{\gamma} \bigr]  \label{stretch}  
\end{equation}   
where the values of the exponents $\gamma$ and $\gamma_x$ are found the same; 
and that iii) the return intervals themself are long-term correlated, with 
an exponent $\gamma'$ close to the exponent $\gamma$ of the original records 
\cite{bunde_physa2003}. The result i) was obtained by Bunde et al.
\cite{bunde_physa2003} by statistical arguments and, as clarified by Altmann 
and Kantz in a very recent paper \cite{kantz_2005}, it can be identified 
with Kac's lemma \cite{kac} in the context of area preserving dynamical 
systems. The results ii) and iii) in Ref. \cite{bunde_physa2003} were obtained
by numerical analysis of long-term correlated Gaussian time series generated 
by the Fourier transform technique and by imposing a power-law decay of the 
power spectrum \cite{makse}. Very recently, it has been realized 
\cite{kantz_2005,bunde_prl2005} that result ii) only applies to linear time 
series (i.e. to series whose properties are completely defined by the power
spectrum and by the probability distribution, regardless of the Fourier 
phases). However, apart from this restriction, the stretched exponential 
distribution of the return intervals of extreme events seems to be a general 
feature in presence of long-term correlations in a time series  
\cite{kantz_2005,bunde_prl2005}. It must be noted that this feature has 
important consequences on the observation of extreme events: in fact it
implies a strong enhancement of the probability of having return periods well 
below $R_q$ and well above $R_q$, in comparison with the 
occurrence of extreme events in an uncorrelated time series. Finally, it 
must be underlined that the distribution in Eq.~(\ref{stretch}) is a 
single parameter function, being $a$ and $b$ in Eq.~(\ref{stretch}) 
functions only of the exponent $\gamma$, as shown in Ref. \cite{kantz_2005}. 
Here, we study the effect on the distribution of return periods of rare 
events of the presence in a time series of finite-term correlations with 
non-exponential decay. Precisely, in the following section we will analyze 
the auto-correlation function and the statistics of the return intervals of 
extreme values of the resistance fluctuations displayed by a resistor 
with granular structure in a nonequilibrium stationary state 
\cite{pen_pre,pen_fnl,pen_physa,pen_ng_fn04,pen_prb}. In particular we will 
consider time series characterized by a stretched exponential decay of 
the auto-correlation function, thus a behavior intermediate between an 
exponential and a power law decay. A situation which can occur in systems 
which are approaching criticality \cite{kotz,sornette}. 

\section{Method and Results}  
We analyze time series of resistance fluctuations of a thin resistor
with granular structure in nonequilibrium stationary states in contact
with a thermal bath at temperature $T_0$ and biased by an external
current $I$. More precisely, we indicate with $R$ the resistance of the
resistor, with $<R>$ its average value and with $\sigma$ the root-mean-square 
deviation from the average and we analyze the normalized
signal: $x(t) \equiv (R(t)-<R>)/\sigma$, with zero average and unit variance.
The $R(t)$ series is calculated by using the stationary and biased
resistor network (SBRN) model introduced in 1999 by Pennetta, Tref\'an and
Reggiani \cite{pen_upon99} and developed in Refs.
\cite{pen_pre,pen_fnl,pen_physa,pen_ng_fn04}. This model describes a thin
film with granular structure as a resistor network in a stationary 
state determined by the competition between two stochastic processes,
breaking and recovery of the elementary resistors. Both processes are
thermally activated and biased by the external current. The
network resistance and its fluctuations are calculated by Monte Carlo 
simulations. All the details about the model and its results are reported
in Refs. \cite{pen_pre,pen_fnl,pen_physa,pen_ng_fn04,pen_prb}. Within this
model, the level of intrinsic disorder in the network (average fraction of
broken resistors in the vanishing current limit \cite{pen_prl_stat}) is
controlled by a characteristic parameter: $\lambda \equiv (E_D -E_R)/k_B T_0$,
where $E_D$ and $E_R$ are the activation energies respectively of the 
breaking and recovery processes. It holds the relationship: 
$\lambda_{min}<\lambda<\lambda_{max}$, where $\lambda_{max}$ 
corresponds to an homogeneous resistor (perfect network) and 
$\lambda=\lambda_{min} \approx 0$ to the maximum level of intrinsic 
disorder compatible with a stationary state of the network (stationary 
resistance fluctuations)
\cite{pen_fnl,pen_physa,pen_ng_fn04,pen_prb,pen_prl_stat}.
In addition to this intrinsic disorder, a disorder biased by the current $I$ 
is also present in the network. As a consequence, for a given value
of $\lambda$, and for a network of given size, nonequilibrium stationary
states of the resistor exist only for $I\leq I_B$ (breakdown threshold). 
When $I>I_B$ the resistor undergoes an electrical breakdown, associated
with an irreversible divergence of its resistance 
\cite{pen_pre,pen_fnl,pen_physa,pen_prb}. For a generic value of
$\lambda$ this breakdown corresponds to a first order transition 
\cite{pen_fnl,pen_physa,pen_ng_fn04}. However, for decreasing $\lambda$
values, when $\lambda \rightarrow \lambda_{min}$, the system becomes more
and more close to its critical point \cite{pen_fnl,pen_physa,pen_ng_fn04}. 
We note that the SBRN model provides a good description of many features
associated with nonequilibrium stationary states and with the electrical
instability of composite and granular materials 
\cite{pen_pre,pen_fnl,pen_physa}, including the electromigration damage 
of metallic lines \cite{pen_prb}. 

\begin{figure}  
\begin{center}
 \resizebox{22pc}{!}{\includegraphics{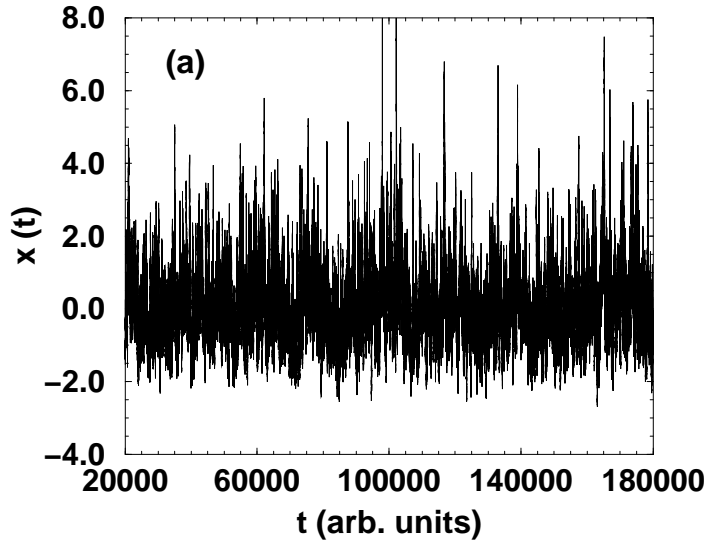}}     
  \caption{The $x(t)$ signal (resistance fluctuations calculated by 
  the SBRN model) as a function of the time expressed in simulation time 
  steps. Only a small portion of the records are shown. The data have been 
  normalized to provide a zero average and a unit variance for $x(t)$.}
\end{center}   
\end{figure}  

\begin{figure}  
\begin{center}
 \resizebox{22pc}{!}{\includegraphics{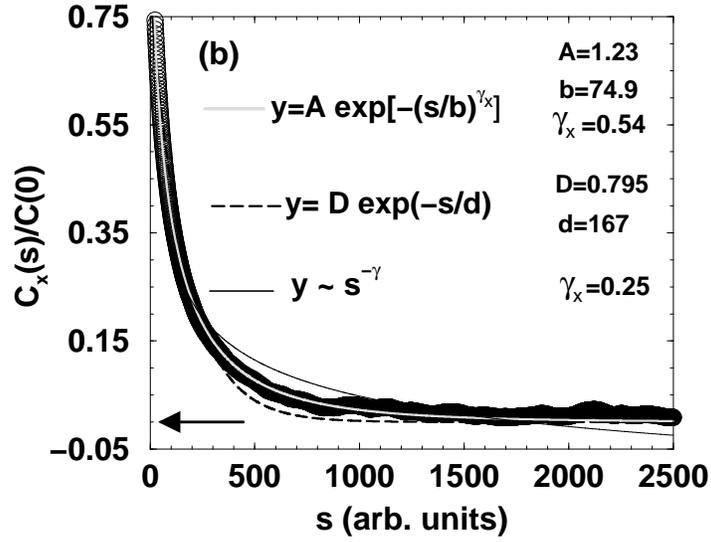}}       
  \caption{The black circles display the auto-correlation function, 
  $C_{x}(s)$, of the signal in Fig. 1. $C_x$ has been normalized to 
  $C_{x}(s=0)$. The black arrow at the bottom points out the origin on 
  the y-axis. The grey curve represents the best-fit with a stretched 
  exponential. The black curves, dashed and solid, show respectively the 
  best-fit with an exponential and a power-law. The figure also reports the 
  values of the best-fit parameters.}
\end{center}   
\end{figure}  

\begin{figure}
 \begin{center} 
 \resizebox{22pc}{!}{\includegraphics{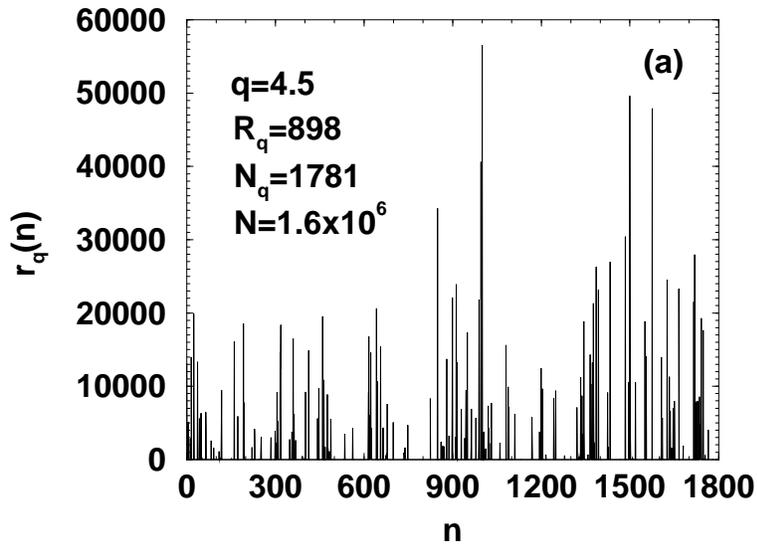}}    
 \caption{Return intervals of an extreme value above the threshold 
 $q=4.5$ (in units of root-mean square deviation) for the time series in 
 Fig. 1. The series consists of $1.6 \times 10^6$ records. The values over 
 the threshold are only $N_q=1781$ and the mean return period is $R_q=898$
 (in simulation time-step units).}   
\end{center}
\end{figure}   

\begin{figure}
 \begin{center} 
 \resizebox{22pc}{!}{\includegraphics{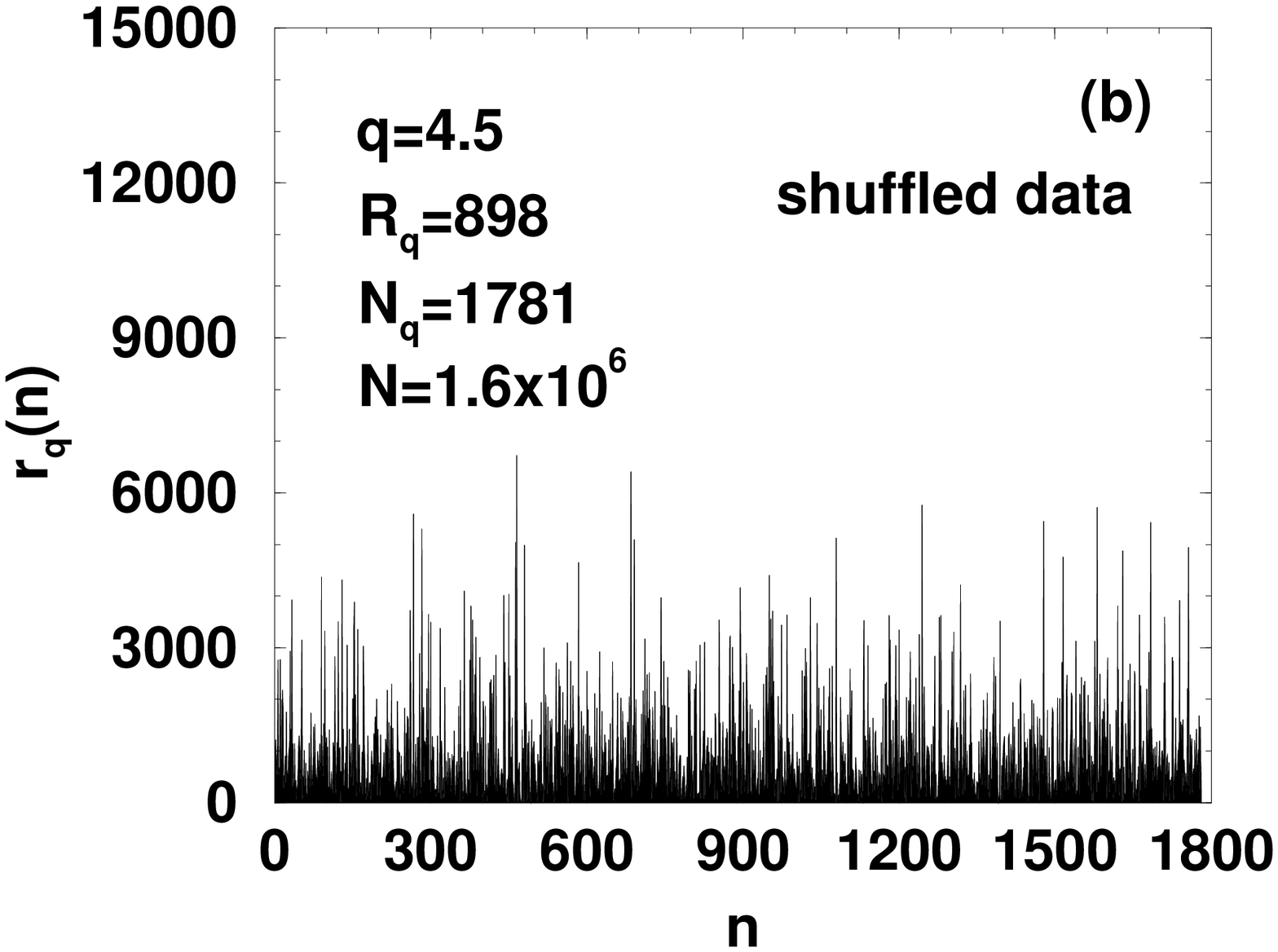}}  
 \caption{The same as Fig. 3 but obtained by shuffling the data 
 (see the text).}   
\end{center}
\end{figure}   

\begin{figure}  
\begin{center}
\resizebox{22pc}{!}{\includegraphics{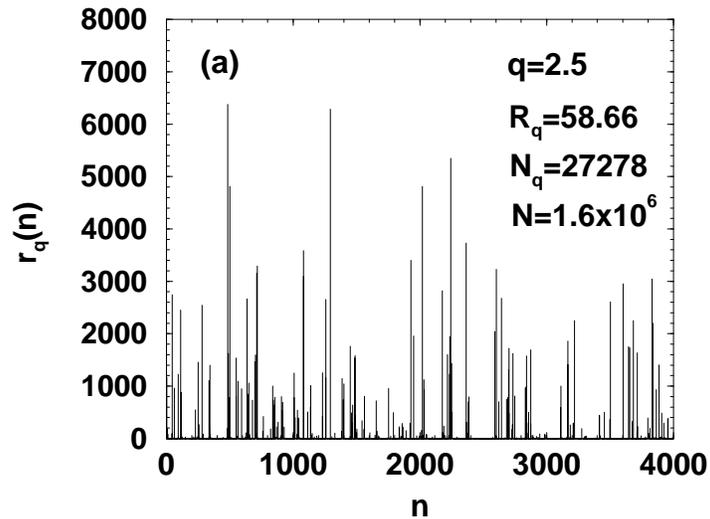}}  
 \caption{Return intervals of an extreme value above the threshold 
 $q=2.5$ (in units of root-mean square deviation) for the same time series 
 considered in the previous figures. The values over the threshold 
 now are $N_q=27268$ and the mean return period is $R_q=58.66$ (in simulation 
 time-step units).}   
\end{center}
\end{figure} 

\begin{figure}  
\begin{center}
 \resizebox{22pc}{!}{\includegraphics{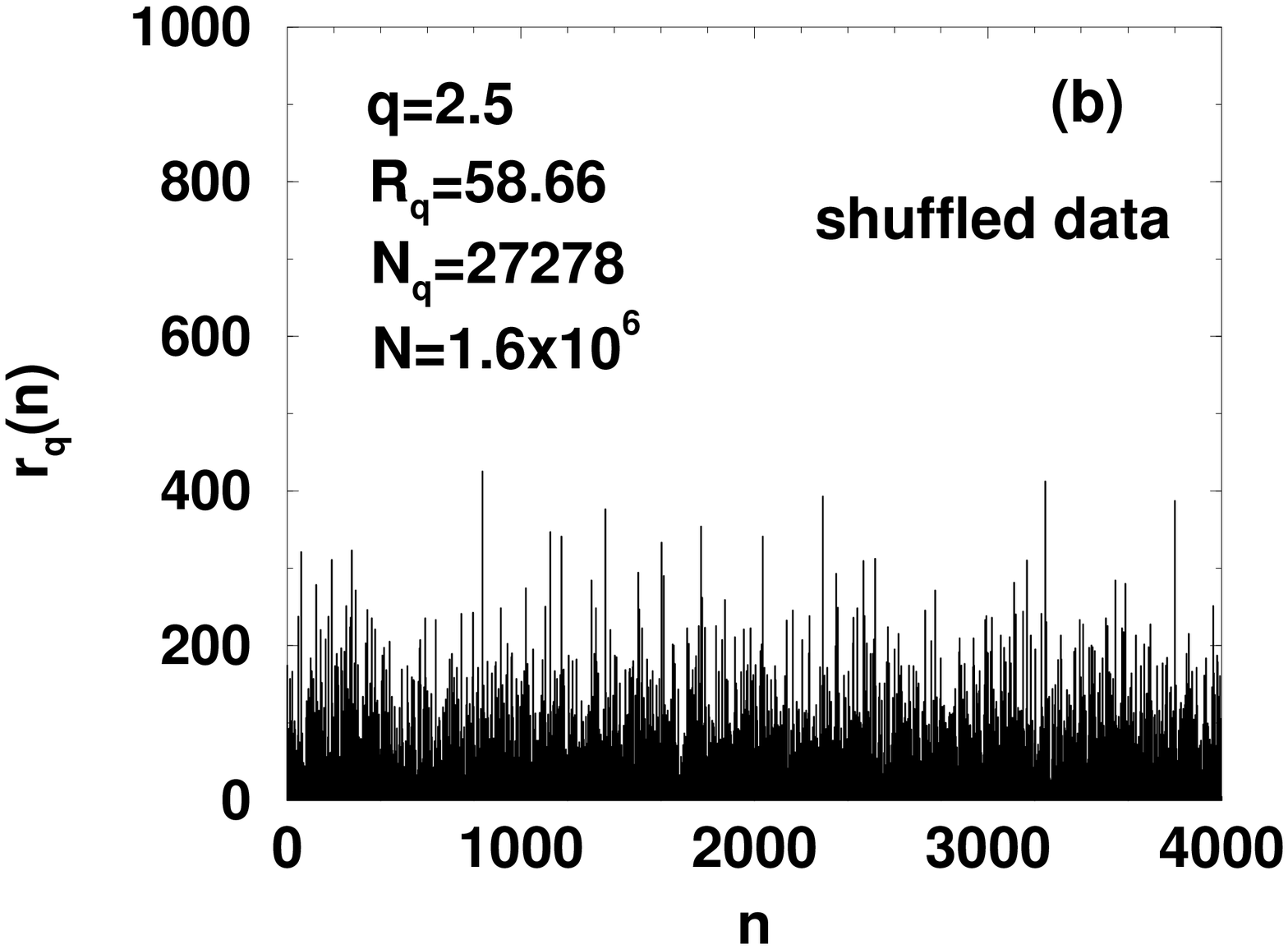}}  
 \caption{The same as Fig. 5 but obtained by shuffling the data 
 (see the text).}   
\end{center}
\end{figure} 

\begin{figure}  
\begin{center}
 \includegraphics[height=.30\textheight]{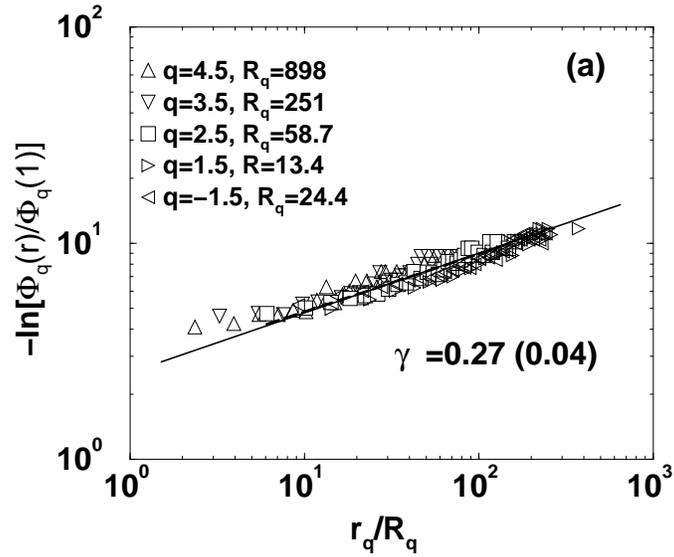}  
 \caption{Double-logarithmic plot of the probability density $P_q(r)$ of 
 the return intervals distribution as a function of $r/R_q$ for different 
 quantiles $q$. The probability densities have been normalized 
 to $P_q(1)$. The slope of the straight line shown in the figure gives the 
 value of the exponent $\gamma$ of a stretched exponential function 
 fitting $P_q(r)$.}
\end{center}   
\end{figure}

 \begin{figure}  
\begin{center}
 \includegraphics[height=.30\textheight]{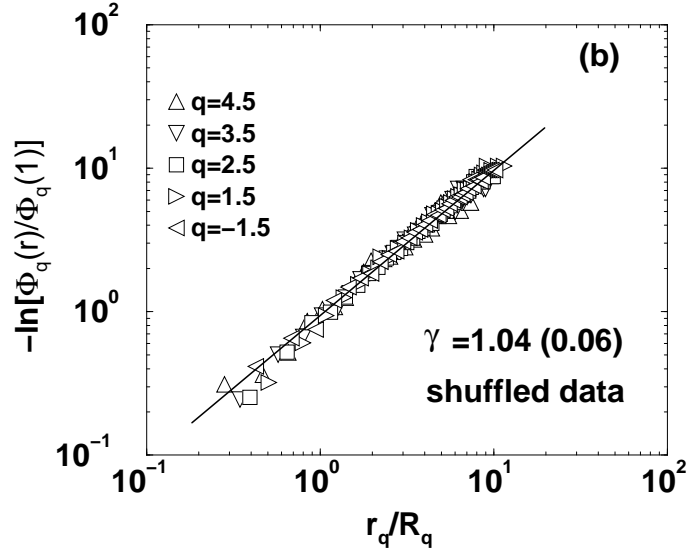}  
 \caption{The same as Fig. 7 but obtained by shuffling the 
 data.}
\end{center}   
\end{figure}

Long $x(t)$ time series (typically made of $1\div 2 \times 10^6$ records)
have been generated and analyzed for different values of $\lambda$, of the
external current and of the network size. The analysis has been performed
by calculating the auto-correlation function $C_x$ and the PDF of the
$x$ records, the return intervals $r_q$ of the extreme values for
different threshold $q$ and their distribution $P_q(r)$. The values of
$q$ are expressed in root-mean-square deviation units. At small
$\lambda$ values (high level of intrinsic disorder), we have found that
the auto-correlation function displays a non-exponential (but non-power-law) 
decay. This behavior is different from that displayed at high
$\lambda$ values (low level of intrinsic disorder), where $C_x$ exhibits
an exponential decay (consistent with the Lorentzian power spectrum
reported in previous works \cite{pen_pre,pen_fnl,pen_ng_fn04}). 
Here we will focus on the case of non-exponential and non-power-law decay
of correlations, a situation typical of systems which are approaching 
criticality, and we will show the results obtained by taking $\lambda=0.33$ 
for a network of size $125 \times 125$, biased by a current $I=0.011$ A. 

Figure 1 displays a portion of the $x(t)$ time series (the total
number of records in $x(t)$ is $1.6 \times 10^6$). The figure clearly shows
a strong non-Gaussianity of the resistance fluctuations. Actually, with
the above specified choice of the parameters, the PDF of the
resistance fluctuations is well described by the Bramwell-Holdworth-Pinton
\cite{bramwell_nat} distribution, as discussed in Refs. 
\cite{pen_physa,pen_ng_fn04}. Figure 2 reports the auto-correlation 
function which significantly deviates from a single exponential or a
power-law behavior. We have found that $C_x$ is well fitted by a
stretched exponential function: $C_x(s)=A\ $ exp$[(-s/b)^\gamma)]$,
with the following values of the fitting parameters: $A=1.23$, $b=74.9$
and $\gamma=0.54$. We have also considered many other functions for the
best-fit of the $C_x$ data. However, we have found that the stretched
exponential function optimizes the best-fit procedure with the minimum
numbers of fitting parameters. We note that this expression of $C_x(s)$
recovers the simple exponential dependence for $\gamma=1$ while it
provides a constant behavior for $\gamma=0$. Moreover, it provides a
finite value of the average correlation time $\tau$.
Figure 3 shows the return intervals of the extreme values above 
the threshold $q=4.5$. For comparison, in Fig. 4 we have reported the
return intervals obtained for the same quantile by random shuffling the
records of $x(t)$ (an operation which leaves unchanged the PDF of the x-data).
The different scales of Fig. 3 and 4 must be underlined. Similarly to
the results of Bunde et al. \cite{bunde_prl2005} concerning
long-term correlated records, Fig. 3 shows a strong clustering of
the extreme events: sequences of very short return intervals follow
sequences of typically long intervals. This strong clustering is present
even if the $x$ records are not long-term correlated while they are
characterized by a finite correlation time. A situation completely
different from that reported in Fig. 4, which corresponds to fully 
uncorrelated records. The clustering of extreme values persists also by
lowering the threshold. This is displayed in Fig. 5 which shows the
return intervals obtained for $q=2.5$. Again, Fig. 6 reports the
returns intervals calculated by shuffling the x-data.

The probability density function  $P_q(r)$ of the return intervals of 
extreme values is plotted in Fig. 7 in a double-logarithmic scale 
as a function of $r/R_q$ and for different quantiles $q$ (the probability 
density has been normalized to $P_q(1)$). This representation has been
adopted for convenience: in fact in this plot a stretched exponential function
with exponent $\gamma$ is represented by a straight line, whose slope just
provides the value of $\gamma$. For check and comparison we have reported in 
Fig. 8 the probability density function $P_q(r)$ of the return intervals 
calculated after random shuffling the x-records: in this case $\gamma=1$, i.e.
the distribution of the $r_q$ is exponential, as it must be for a Poisson
process. Thus, Fig. 7 shows that the distribution of return intervals of 
extreme values of $x(t)$ is well described by a stretched exponential and that
the value of the exponent $\gamma$ is independent of the threshold $q$ in a 
large range of $q$-values. Moreover, at least in the case of the data reported
in Figs. 1-8, the value of the exponent $\gamma$ of the return interval 
distribution is just $1/2$ of the value of  $\gamma_x$, the exponent of the 
auto-correlation function. An analysis performed on other sets of data, 
obtained for high disordered networks of different size $L$ and/or biased by 
different current density $I/L$, seems to confirm this last result. 
However, further investigations are necessary to establish the possible 
relationship between the two exponents.    
  
\section{Conclusions and Open Questions}  
We have studied the distribution of return intervals of extreme events
in time series with finite-term correlations. Precisely, we have analyzed 
the auto-correlation function and the distribution of the return intervals of 
extreme values of the resistance fluctuations displayed by a resistor with 
granular structure in nonequilibrium stationary states. The resistance 
fluctuations were calculated by Monte Carlo simulations using the SBRN model 
\cite{pen_pre,pen_fnl,pen_physa,pen_ng_fn04,pen_prb}. We have found that for 
highly disordered networks, when the auto-correlation function displays a 
non-exponential (precisely, a stretched exponential) decay and the 
resistance fluctuations are characterized by a finite correlation time, 
the distribution of return intervals of the extreme values is well described 
by a stretched exponential with exponent $\gamma$ largely independent of the 
threshold $q$. Therefore our results show that the stretched exponential 
distribution \cite{sornette} describes the distribution of return intervals 
of extreme events not only in the case of long-term correlated time series 
\cite{kantz_2005,bunde_physa2003,bunde_prl2005} but also when the records are 
characterized by finite-term correlations, with non-exponential decay of the 
auto-correlation function. Many open questions arise from this study and we 
limit to mention the following ones: i) why is the distribution of return 
intervals of extreme values a stretched exponential? what is the basic reason 
for this behavior which, as we have found, is not only limited to long-term 
correlated records \cite{kantz_2005,bunde_physa2003,bunde_prl2005} but is 
present also when non-exponential, finite term correlations are present? ii) 
what is the relationship between the exponents $\gamma$ and $\gamma_x$? 
The correlations among the $r_q$ have been studied in Refs. 
\cite{kantz_2005,bunde_physa2003,bunde_prl2005} in the case of long-term 
correlated processes, thus we can ask iii) what kind of correlations exist 
among the $r_q$ in the case of a finite-term, non-exponentially correlated 
processes ? 

\vspace*{0.2cm}  
This work has been partially supported by the SPOT NOSED project 
IST-2001-38899  of E.C. and by the cofin-03 project "Modelli e misure in   
nanostrutture" financed by Italian MIUR. The authors thank S. Ruffo 
(University of Florence, Italy), P. Olla (ISAC-CNR, Lecce, Italy) and
G. Salvadori (University of Lecce, Italy) for helpful discussions.   
  
\bibliography{upon_event}  
  
\end{document}